\documentclass[useAMS,usenatbib]{mn2e}
\usepackage{psfig}
\usepackage{url}
\usepackage[dvips]{graphicx}
\def\mnras{{MNRAS}}
\def\apj{{ApJ}}
\def\aj{{AJ}}
\def\aap{{A\&A}}
\def\apjl{{ApJL}}

\def\nat{{Nature}}

\def\pasa{{PASA}}

\def\xmm{{\sl XMM-Newton}}

\def\mcg6{{MCG-6-30-15}}

\def\msun{$M_{\odot}$}
\def\tb{{$T_{B}$}}

\def\me{{$\dot{m}_{E}$}}
\def\nub{{$\nu_{B}$}}

\def\nxs{{$\sigma^2_{\rm NXS}$}}
\def\ltsim{\mathrel{\hbox{\rlap{\hbox{\lower4pt\hbox{$\sim$}}}\hbox{$<$}}}}
\def\gtsim{\mathrel{\hbox{\rlap{\hbox{\lower4pt\hbox{$\sim$}}}\hbox{$>$}}}}

\begin{document}

\title[Black hole mass and AGN X-ray Variability Amplitude]{Short
  Timescale AGN X-ray Variability with EXOSAT: Black hole mass and Normalised Variability Amplitude}

\author[M$\rm^{c}$Hardy, I.M.]
{ I.M. M$\rm^{c}$Hardy$^{1}$\\
$^{1}$ Department of Physics and Astronomy, The University, Southampton 
SO17 1BJ
}

\maketitle

\begin{abstract}
The old EXOSAT medium energy measurements of high frequency (HF) AGN
power spectral normalisation are re-examined in the light of accurate
black hole mass determinations which were not available when these
data were first published \citep{green93}. It is found that the
normalised variability amplitude (NVA), measured directly from the
power spectrum, is proportional to $M^{\beta}$ where $\beta
\sim-0.54\pm 0.08$. As NVA is the square root of the power, these
observations show that the normalisation of the HF power spectrum for
this sample of AGN varies very close to inversely with black hole
mass.  Almost the same value of $\beta$ is obtained whether the quasar
3C273 is included in the sample or not, suggesting that the same
process that drives X-ray variability in Seyfert galaxies applies also
to 3C273.  These observations support the work of
\cite{gierlinski08_mass} who show that an almost exactly linear
anticorrelation is required if the normalisations of the HF power
spectra of AGN and X-ray binary systems are to scale similarly.  These
observations are also consistent with a number of studies showing that
the short timescale variance of AGN X-ray lightcurves varies
approximately inversely with mass.

\end{abstract}

\section{Introduction}
\label{sec:intro}
It is now generally agreed that the X-ray variability properties of
$\sim$few solar mass black holes in X-ray binary systems (XRBs) and
supermassive black holes in active galactic nuclei (AGN) are
similar. With a view to providing an independent method for determining
physical parameters such as black hole mass and accretion rate, there
is considerable interest in determining how these parameters might
scale with characteristic observable X-ray variability
parameters. 

X-ray variability can be quantified via the
power spectral density (PSD) and, to first order, the X-ray PSDs of
AGN are similar to those of XRBs \citep{mch88}, particularly to XRBs in
the high-soft state \cite[e.g.][]{uttley02,mch04}. In this state 
the power, $P$, at frequency, $\nu$, is given by $P(\nu) \propto
\nu^{-\alpha}$ where $\alpha \sim 1$ at low frequencies bending,
above a frequency $\nu_{B}$, to a slope $\alpha \geq 2$.
Many authors have shown that the timescale $T_{B}$,
corresponding to the frequency $\nu_{B}$, scales approximately
linearly with mass \cite[e.g.][]{mch88, edelson99, uttley02,
  markowitz03_psd,mch04, mch06,kelly11,gonzalez-martin12} although the
relationship shows considerable scatter. However much of the scatter
can be explained by an inverse scaling of \tb\, with accretion rate
\me\, \cite[where \me\, is the accretion rate in units of the
Eddington accretion rate,][]{mch04,uttleymch05,mch06,kording07a}.

The other variability parameter which has attracted attention is the
normalisation of the high frequency (HF) PSD. This normalisation can
be defined in a number of ways. For example, by fitting a power law to
the HF PSDs, after subtraction of the Poisson noise contribution,
\cite{mch88} measured the power at a fixed frequency ($2 \times
10^{-4}$Hz).  The normalised variability amplitude, NVA, was then
defined as the square root of the power, which was measured in units
of (counts s$^{-1}$)$^{2}$ Hz$^{-1}$, divided by the average count
rate. The NVAs for a sample of AGN observed by EXOSAT were therefore
derived. Although black hole mass was assumed to be a driving
parameter, in 1988 very few AGN black hole masses were available so
the NVAs were plotted against luminosity as a proxy, showing a strong
anticorrelation.

\cite{hayashida98}, using GINGA observations, measured the timescale
at which the PSD crossed a particular power level ($10^{-3}$, in
rms$^{2}$ Hz$^{-1}$ units). Scaling this timescale with mass
from Cyg X-1, they estimated masses for 8 AGN,
although they noted that these masses were one or two orders
of magnitude lower than masses derived by other methods.
\cite{gierlinski08_mass} defined an HF PSD normalisation, $C_{m}$,
which is very similar to the NVA of \cite{mch88} although, as most of
their measurements were of XRBs, \citeauthor{gierlinski08_mass}
defined $P(\nu)$ at the higher frequency of 1Hz. 
They also assumed $\alpha=2$ whereas \cite{mch88} and \cite{green93}
measured $\alpha$.
Within the XRB sample of \citeauthor{gierlinski08_mass} there is no
correlation of $C_{m}$ with mass (their Fig.7a), although the range of
masses is small and mass uncertainties large. Within their sample of
AGN (their Fig.7c) there is an approximate inverse correlation of
$C_{m}$ with mass, strengthened if NGC4395 is included
(Fig.7b). However the scatter is too large to define the relationship
precisely within that sample.  However if it is assumed that AGN and
XRBs follow the same scaling relationship then
\cite{gierlinski08_mass} find that $C_{m} \propto M^{-0.98\pm 0.01}$.
\cite{kelly11} fit a mixed Ornstein-Uhlenbeck model to Cyg X-1 and to
AGN PSDs. In this model the HF PSD normalisation is
defined by $\zeta$, the fractional amplitude of the driving noise
field.  The PSD slope is again fixed at $\alpha=2$ with $P(\nu)
\propto \left( \frac{\zeta}{\nu} \right)^{2}$.  For a sample of 10 AGN
they find a strong correlation with $\zeta \propto M^{-0.79 \pm0.22}$
(90 per cent confidence).
However such a relationship would not extrapolate to the XRB
observations and so it is important to investigate the relationship
between AGN HF PSD normalisation and mass using other datasets.

Although less direct, the PSD normalisation can be estimated from the
normalised excess variance of the lightcurves, $\sigma^2_{\rm NXS}$
\cite[e.g.][]{nandra97a}. A number of authors
\cite[e.g.][]{papadakis04,nikolajuk04,oneill05,zhou10,ponti12} have
found an approximate inverse scaling of $\sigma^2_{\rm NXS}$ with
mass. If all of the frequencies sampled by the lightcurves lie above \nub,
and the HF PSD slope is known, then \nxs\, gives a reasonable estimate of the
HF PSD normalisation, although \cite{vaughan03_variability}
show that \nxs\, is a noisy quantity when the PSD is steep.
If \nub\, is not known and lies within
the frequencies sampled, or if the low frequency PSD
slope is unknown, then \nxs\, is a less good measure of HF PSD
normalisation.

The best way, therefore, to determine whether HF PSD normalisation
varies with mass is to derive HF PSD normalisations directly from the
PSDs. The HF PSD is best determined from continuous
long observations which suffer less from distortion by the window
function of the sampling pattern than do observations which are split
into many segments. Thus \xmm, which allows continuous observations of
up to 130ksec, provides a better measurement of the HF PSD
\cite[e.g.][and many others]{gonzalez-martin12,mch05a} than do
satellites in low earth orbits with maximum continuous observation
lengths of $\sim4000$s, i.e.  GINGA \cite[e.g.][]{hayashida98}, ASCA
\cite[e.g.][]{oneill05} or RXTE.

EXOSAT was also useful as it allowed continous observations of up to
280ksec. The EXOSAT low energy (LE) imaging telescope is not
comparable in sensitivity at low energies (0.1-2 keV) to \xmm. However
the EXOSAT Medium Energy proportional counter (ME) is comparable at
the higher energies (approximately 1-9 keV) where, also, the effects
of absorption which may affect measurements of X-ray flux variations,
are less.  In addition, the ME has made long observations of some
bright AGN (eg 3C273, CenA, NGC5506) for which long observations by
\xmm\, do not yet exist. In this paper the EXOSAT ME observations of AGN
are re-examined to determine whether they reveal any scaling of HF PSD
normalisation with mass.

\section{EXOSAT Observations}

\subsection{The EXOSAT ME Sample}

\cite{mch88} provided initial NVA measurements for a sample
of AGN observed by EXOSAT, but a more precise analysis of all EXOSAT ME
AGN observations longer than 20ksec (32 AGN in total) was carried out
by \cite{green93}. \citeauthor{green93} removed the small window
artefacts from EXOSAT PSDs using their own 1D version of the `CLEAN'
algorithm \citep{roberts87} and hence derived NVAs. Errors
were determined from simulations. \citeauthor{green93} confirmed the
inverse scaling of NVA with luminosity, suggesting that the lower NVAs
arose in larger mass objects from a larger emitting region, and also
found a positive correlation of NVA with photon energy index. 

Where variability was not detected, ie the hypothesis that the
lightcurve was constant could not be rejected at the 95 per cent
confidence level at least 90 per cent of the time,
\citeauthor{green93} provide an NVA upper limit, deduced from the
photon counting noise level of the lowest noise observation.  The
reader is referred to \citeauthor{green93} for full details.
\begin{table}
\begin{tabular}{lrcc}
Name & Black Hole Mass & Ref & NVA \\
     & ($\times 10^{6} \,M_{\odot}$) &       &               \\
& & &  \\   
     3C120 &   55$^{+ 31}_{-22}$       &  1 & $1.59\pm0.82$  \\  
   NGC3227 &   15$^{+  5}_{ -8}$       & 2 & $1.27\pm0.60$  \\   
   NGC3783 &   29.8$^{+  5.4}_{ -5.4}$  & 1 & $1.59\pm1.13$  \\   
   NGC4051 &    1.73$^{+0.55}_{ -0.52}$ & 3 & $5.78\pm2.08$  \\   
   NGC4151 &   45$^{+  5}_{ -5}$       & 4 & $0.76\pm0.27$  \\   
     3C273 &  890$^{+190}_{-190}$      &  1 & $0.20\pm0.08$  \\   
   NGC4593 &    9.8$^{+2.1}_{ -2.1}$   & 5 & $3.13\pm1.15$  \\   
MCG-6-30-15&    4.5$^{+  1.5}_{ -1.0}$  & 6 & $3.87\pm0.92$  \\   
   NGC5506 &    7.0$^{+  3.5}_{ -3.5}$  & 7 & $1.92\pm0.35$  \\   
   NGC7314 &    0.87$^{+0.45}_{ -0.45}$ & 7 & $6.68\pm2.56$  \\  
 MCG+8-11-11 &    --                 & -- & $0.45\pm0.16$    
\end{tabular}
\caption{NVAs from  \protect\cite{green93}
for AGN with detected variability in EXOSAT ME observations.
Black hole masses are from 1 --
  \protect\cite{peterson04}, 2 --\protect\cite{davies06}, 
3 -- \protect\cite{denney10}, 4 --\protect\cite{onken07}, 
5 --\protect\cite{denney06}, 
    6 -- \protect\cite{mch05a}, 7 -- \protect\cite{gu06}. }
\label{tab:detect}
\end{table}
\begin{table}
\begin{tabular}{lrcc}
Name &Black Hole Mass & Ref & NVA\\
     & ($\times 10^{6} \,M_{\odot}$) &       &  limit        \\
& & &  \\   
      MKN335 &    14.2$^{+  3.7}_{ -3.7}$   &  1 &    4.30  \\   
      IIIZw2 &           --               & -- &    2.90  \\  
    FAIRALL9 &   255$^{+ 56}_{-56}$         &  1 &    3.30  \\   
      NGC526 &        --                  &  -- &    2.90  \\  
     Mkn1040 &        --                  &  -- &    3.50  \\  
    0241+622 &        --                  &  -- &    2.90  \\  
      Akn120 &   150$^{+ 19}_{-19}$         &  1 &    1.60  \\  
     NGC2110 &   200$^{+100}_{-100}$        & 8 &    5.20  \\   
  3A0557-383 &          --                & -- &    1.80  \\  
     NGC2992 &    52$^{+ 25}_{-25}$         & 8 &    1.41  \\   
     NGC3516 &    31.7$^{+  2.8}_{ -4.2}$    &  3 &    3.90  \\   
        CENA &    50$^{+ 10}_{-10}$         & 9 &    0.60  \\   
      IC4329 &   217$^{+180}_{-105}$        &  10 &    0.70  \\   
     NGC5548 &    65.4$^{+  2.6}_{ -2.5}$    &  11 &    1.53  \\  
   E1821+643 &              --            & -- &    3.40  \\  
       3C382 &              --            & -- &    1.40  \\  
     3C390.3 &   287$^{+ 64}_{-64}$         &  1 &    3.40  \\  
  MR2251-179 &   100$^{+ 50}_{-50}$         & 12 &    2.40  \\   
     NGC7469 &    12.2$^{+  1.4}_{ -1.4}$    &  1 &    2.00  \\   
  MCG2-58-22 &              --            &  -- &    2.30    
\end{tabular}
\caption{Upper limits to NVAs from \protect\citeauthor{green93}
for AGN with no detectable variability in EXOSAT ME observations.
  Mass references are as in Table 1 with
8 -- \protect\cite{woo02a}  9 -- \protect\cite{neumayer10}  
10 -- \protect\cite{markowitz09}
11 -- \protect\cite{bentz07}  12 -- \protect\cite{zhou05}
  For CenA the mass given is the average of the values derived
  from $H_{2}$ kinematics ($4.5 \times 10^{7}$ \msun) and from stellar
  kinematics ($5.5 \times 10^{7}$ \msun), taking the 1$\sigma$ value
  of the larger, stellar kinematics, error.
 Where masses are not from reverberation mapping and
  no precise mass error is quoted,
ie for NGC2110, NGC2992, MR2251-179, a 50 per cent error is assumed. }
\label{tab:undetect}
\end{table}

\subsection{NVA and black hole mass measurements}

In Table~\ref{tab:detect} the NVAs where variability was detected are
listed together with the most accurate black hole masses available.
Masses derived from stellar dynamical observations (NGC3227 and
NGC4151), which are not subject to the uncertainty regarding the
`f-factor' used to convert dynamical products from reverberation
observations to black hole masses, are taken preferentially. As
dynamical measurements are rare, reverberation masses are next taken
(3C120, NGC3783, NGC4051, 3C273 and NGC4593; see the caption to Table
1). For MCG-6-30-15 the mass estimated by \cite{mch05a} is used. This
mass is the mean of the mass derived from the width of the stellar
absorption lines and from the emission line width. For NGC5506 and
NGC7314 the mass is derived from the stellar absorption line widths of
98 and 60 km s$^{-1}$ respectively \citep{gu06} using, as for
MCG-6-30-15, the $M-\sigma$ relationships of \cite{merritt01} and
\cite{tremaine02}. The latter, particularly, has been shown by
\cite{greene04} to provide a good mass estimate for low mass
systems. The mean of the masses derived from these two relationships
is taken and a 50 per cent uncertainty is assumed.  No reliable mass
estimator could be found for MCG+8-11-11 and so this AGN is not
considered further. In Table~\ref{tab:undetect} the NVA upper limits
for sources where variability was not detected are listed.  Black hole
masses are also given.

\subsection{Quantifying the NVA-Mass Relationship}

A simple maximum likelihood (ML) analysis, using the function
within the {\sc QDP} plotting package which takes only the errors on
NVA, was applied to the AGN with detected variability (except for
MCG+8-11-11). This analysis shows that the relationship  $NVA \propto
M^{\beta}$, with
$\beta = -0.52 \pm 0.11$ (90 per cent confidence) is a good
description of these data. The {\sc QDP} ML W-Var statistic, which is
similar to $\chi^{2}$, is 6.1 for 8 d.o.f., indicating
that the errors are overestimated. To determine whether $\beta$
varies with the method of fitting, a number of other
methods were examined, particularly {\sc fitexy} \citep{press92} and
Bayesian regression \citep{kelly07}.

\begin{figure}
\psfig{figure=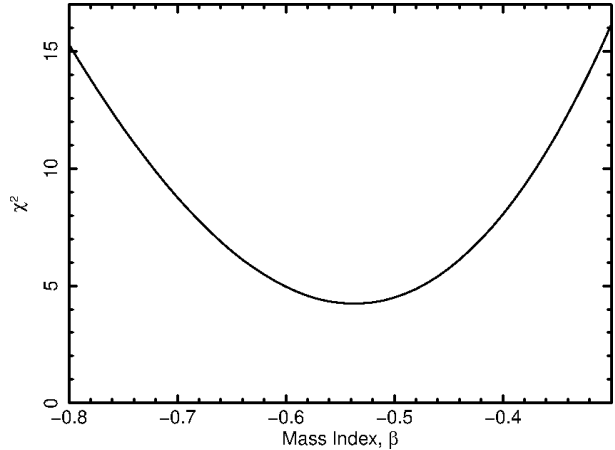,width=80mm,angle=0}
\caption{$\chi^{2}$ vs mass index, $\beta$, for the fit to a 
relationship of the form NVA $\propto M^{\beta}$ for all
10 AGN with detectable
variability and well determined masses, i.e. 
those marked with black crosses in Fig.~\protect\ref{fig:nva}.
The minimum value of $\chi^{2}$ corresponds to $\beta=-0.54$.
}
\label{fig:slopes}
\end{figure}

\begin{figure}
\psfig{figure=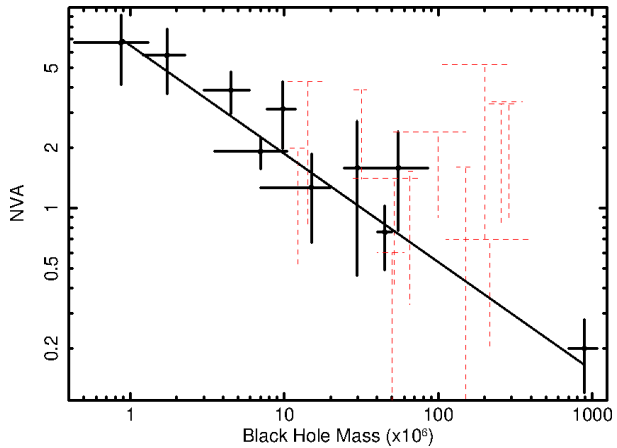,width=80mm,angle=0}
\caption{NVA vs black hole mass.  The line shows NVA $ \propto
  M^{-0.54}$. Although not included in the fit, the
  upper limits (red dashed lines) are consistent with this fit.}
\label{fig:nva}
\end{figure}

In the {\sc fitexy} method, errors in both directions are included. The
best value and uncertainty in $\beta$ were determined by measuring the
minimum value of $\chi^{2}$ as a function of $\beta$, with free
normalisation  (Fig.~\ref{fig:slopes}). The lowest
overall $\chi^{2}$ of 4.24 corresponds to $\beta= -0.54$. 
For 8 d.o.f., this $\chi^{2}$  value confirms that the
errors are over-estimates. The 1$\sigma$ (ie
$\delta \chi^{2}$=1) uncertainty is 0.08. The 90 per cent confidence
uncertainty (ie $\delta \chi^{2}$=2.7) is 0.12.  The Bayesian
regression analysis of \cite{kelly07}, as implemented within IDL gives
an almost identical value of $\beta= -0.53$. The standard deviation
of 0.10 is slightly larger, probably because Bayesian analysis takes
account of the fact that the observed AGN are just a random
sample drawn from the parent distribution, a consideration
addressed here in Section~\ref{sec:sample}. 
Thus the results do not depend on
the fitting method.  The best fit from the {\sc fitexy} method is shown
in Fig.~\ref{fig:nva}. Unless stated otherwise, {\sc fitexy} results
will be quoted hereafter. The NVA upper limits are not included in this fit
but are consistent with the fit.

In principle the NVA upper limits can be used to refine the value of
$\beta$. However the result depends on how the
`upper limits' are interpreted and what probability density functions (PDFs)
are assigned to them, eg whether they are hard limits above
which there is no probability of finding the datum, or
limits with an associated measurement error. Eg the methodology of
\cite{kelly07}, as implemented within IDL, treating the limits as
hard, gives $\beta= -0.59 \pm 0.09$. Allowing some error in the limit
allows slightly flatter values.
However as it is not clear how the limits of \citeauthor{green93}
should be interpreted, it is concluded only that the upper limits are
consistent with values of $\beta$ derived from detections,
but not favouring values of $\beta$ much flatter than -0.5.

For almost all of the present sample it is now known that
\nub\, is well below $2 \times 10^{-4}$Hz \cite[e.g. see][]{mch06} and
so the NVAs reported by \citeauthor{green93} are a good measurement of the
normalisation of the HF PSD. In addition, where PSD slopes could be
reasonably measured, $\alpha$ is mostly close to 2, as expected
for the HF part of the PSD. However for NGC4051 $\nu_{B}=8 \times
10^{-5}$Hz \citep{mch04} and the EXOSAT observation length (207ks) is
long enough that a single power law fit would be noticeably flattened
by the PSD values below \nub, where $\alpha \sim
1.1$. Re-examination of the EXOSAT PSD shows that the NVA could be
underestimated by up to 50 per cent. With a corrected NVA, $\beta=-0.57
\pm 0.08$.
Parameterising log(NVA)=$\beta$log(M) +$C$, then here
$C=0.93 \pm 0.11$.

The only other source where a significant underestimation of the NVA
might be expected, based on its low mass, is NGC7314. However the
EXOSAT observation is only 22ks and so the HF PSD is unlikely
to be noticeably distorted. Even allowing for an extreme 50 per cent
underestimate similar to that for NGC4051, and also allowing for the same
underestimate in NGC4051, $\beta$ increases only to $-0.59 \pm 0.07$.

Measurement errors in the covariate, in this case mass, bias the slope
towards zero \cite[e.g.][]{akritas96a}. Although not discussed in
detail here, that bias can be addressed via simulations
where each data point is represented by a 2D PDF. Random selection of
simulated data points from those PDFs gives values of $\beta$ which
are typically flatter by 0.03 to 0.04 than those obtained by {\sc
  fitexy}. As the simulations include the measurement error twice,
i.e. once in the value of the data point itself and once in the
distribution of the PDF, but {\sc fitexy} only includes the error
once, the true values of $\beta$ may be $\sim 0.035$ steeper than
those given above.

\subsubsection{Dependence of NVA on Sample Parameters}
\label{sec:sample}

It is often questioned whether the X-ray emission mechanism in 3C273
is similar to that in Seyfert galaxies, ie thermal Compton scattering
from a non-beamed corona, or whether it is synchrotron or
synchrotron self-Compton emission from a relativistic jet
\cite[e.g.][]{haardt98,mch99b,mch07_273}. 3C273 shows a bend in
its long term PSD which is consistent with arising from the same
variability process as in Seyfert galaxies \citep{mch06_273} but it is
still unclear whether it should be included together with Seyfert
galaxies and non-blazar radio galaxies (3C120), ie the remainder of
Table~\ref{tab:detect}, in any X-ray timing survey.
The fit was therefore repeated without 3C273, providing a
similar slope ($\beta=-0.54 \pm 0.12$ from {\sc fitexy} or $-0.51 \pm
0.17$ from Bayesian regression) whose extrapolation
passes well within the error range for 3C273. These results suggest
that the same process that drives X-ray variability in Seyferts
also drives variability in 3C273 although
the X-ray emission location or emission mechanism need not be
the same in both cases.

With small samples such as that listed in Table~\ref{tab:detect}, the
results can depend on the choice of the values of the measured
variables and/or of the sample content. For example for
MCG-6-30-15 one might take the mass derived only from the stellar
absorption lines \cite[$5.7 \times 10^{6}$ \msun,][]{mch05a}, which
would not alter $\beta$.  Or one might decide to take masses derived
from reverberation for NGC3227 \cite[$7.36_{-1.72}^{+1.62} \times
  10^{6}$ \msun,][]{denney10} and NGC4151 \cite[here identical to the
  value listed in Table~\ref{tab:detect},][] {onken07}. The value of
$\beta$ is again almost unchanged ($\beta=-0.53 \pm 0.07$ or
$\beta=-0.54 \pm 0.012$ if 3C273 is excluded) although the fit, whilst
remaining good, is very slightly worse.  One might even decide to
exclude MCG-6-30-15, NGC5506 and NGC7314 altogether, giving $\beta
=-0.55 \pm 0.10$.

It is finally noted that \cite{graham11} propose a revision of black hole
masses. Using their `classical' (ie all morphological type) $M-\sigma$
relationship, the masses of MCG-6-30-15 and NGC5506 would reduce by
almost exactly a factor 2, and that of NGC7314 by a factor 3.  All masses
based on reverberation mapping would reduce also by a factor 2 due to
their downward revision of the `f-factor'. As these changes are almost
entirely systematic, the various fits for $\beta$ are almost unchanged
with typical values flattening by only $\sim 0.02$ (eg from -0.54 to
-0.52) with, if anything, very slightly reduced dispersion.

The overall conclusion is that $\beta$ is robust to minor changes in
sample parameters with a value close to -0.54.

\section{Discussion}

It is shown here that NVA, defined by \cite{mch88} and derived more
precisely from EXOSAT ME observations by \citeauthor{green93}, correlates
very well, within a sample of AGN alone, with black hole mass. For a
relationship  $NVA\propto M^{-\beta}$ then, even allowing
for different choices of mass or sample content,
$\beta$ remains close to -0.54 with a dispersion of $\sim0.08$. The main
contribution of the present work is thus to show that a close to
inverse linear scaling of HF PSD normalisation with mass,
proposed by \cite{gierlinski08_mass} on the assumption that AGN and
XRBs follow the same scaling, occurs independently within a sample of
AGN alone.

The large scatter within the AGN $C_{m}$ values of
\citeauthor{gierlinski08_mass} may have arisen because their values of
$C_{m}$ were derived from earlier measurements of X-ray excess
variance by satellites with large orbital gaps in their lightcurves,
ie ASCA and RXTE. Variations in accretion rate may perhaps add further
scatter.  Within their sample of XRBs, \citeauthor{gierlinski08_mass} show
that $C_{m}$ is not constant for any given XRB, although they note
that the range of variability with flux is not large. They do,
however, note that the value of $C_{m}$ in the soft state of Cyg X-1
is larger than in the hard state
which may indicate some variation of HF PSD normalisation with
accretion rate or state. However the range of accretion rates within
the sample listed in Table~\ref{tab:detect} is not large. Thus, as
there is no evidence for dispersion in the present value of
$\beta$ over and above that expected from the errors in the mass and
NVA measurements, possible variation of NVA with accretion rate are
not considered further here although they cannot be ruled out.

Finally, the differences between the results of \cite{kelly11} and
those presented here are considered. The equivalent value of the slope
$\beta$ derived by \citeauthor{kelly11} from Bayesian regression is
$-0.79 \pm0.22$ (90 per cent confidence) whereas the value presented
here is $-0.54 \pm 0.08$ (1$\sigma$, or $\pm0.12$ at 90 per cent
confidence), or -0.57 if NGC4051 is corrected.
Possible reasons for the differences in slope include differences in
analysis methods and in sample content.

Regarding analysis methods, for the present
sample all regression methods produce the same value of $\beta$. 
Another analysis difference is that the  NVA values are derived
from simple fits to the HF PSDs whereas the values of $\zeta$ are
derived from fitting a particular model, the mixed
Ornstein-Uhlenbeck model, to the datasets. Although parameters in
multiparameter fits usually interact, there is no obvious reason why
this model should result in larger values of $\beta$. Similarly fixing
the HF PSD slope at $\alpha=2$ (\citeauthor{kelly11,gierlinski08_mass})
compared with measuring $\alpha=2$ (\citeauthor{green93}) and then correcting
the NVA if $\alpha$ is lower than 2 should have little effect.

Large differences in sample content may, however, have a large effect
on the value of $\beta$.  Only 6 of the 10 AGN in each sample
are in common.  \citeauthor{kelly11} include the high accretion rate
NLS1s Mkn766 and Ark564 (whose mass is uncertain). However
these two NLS1s are not included here where the spread in
accretion rate is more limited.  Mkn766
and Ark564 are both low mass AGN and so have large leverage on the
value of $\zeta$. If there is a slight increase of HF PSD
normalisation with accretion rate or `state', as the observations of
Cyg X-1 by \citeauthor{gierlinski08_mass} suggest,
inclusion of these two AGN would steepen $\zeta$.
Also some masses used here differ from those used by
\citeauthor{kelly11}, taken from
\cite{sobolewska09}.  With small samples, such differences
in content can have a significant effect. Further studies to determine
$\beta$ for other samples of AGN are clearly merited.

The present results are also consistent with the work of
\cite{ponti12} who measure the excess variance, $\sigma^2_{\rm NXS}$,
for a sample of AGN observed by \xmm.  They find that $\sigma^2_{\rm
  NXS} \propto M^{-\gamma}$ where $\gamma=1.15\pm 0.12$, which is very
close to the relationship derived here, for the AGN with 80ksec
minimum duration observations or $\gamma=1.32\pm 0.14$ for AGN with a
minimum of 40~ksec duration.

\vspace*{-2mm}
\section{Conclusions}

Using NVA measurements from \cite{green93} it has been shown that,
within a sample of AGN observed by EXOSAT, HF PSD normalisation scales
almost exactly inversely with black hole mass. These observations
support the proposal of \cite{gierlinski08_mass} that HF PSD
normalisation scales exactly inversely from XRBs to AGN. It is also
noted that the quasar 3C273 fits well onto the scaling relationship
derived for Seyfert galaxies, suggesting that the same process which
drives X-ray variability in Seyfert galaxies also drives X-ray
variability in 3C273, even though the emission process or emission
location may be different.

\section*{Acknowledgements}
I thank Dimitrios Emmanoulopoulos and Christian Knigge for extensive
discussions about statistics and regression and I thank Dimitrios for
tuition in the use of Mathematica. I thank Liz Bartlett for advice
regarding IDL and Brandon Kelly for discussions on regression
and for advice as to how to run his Bayesian regression code inside
IDL. I thank the anonymous referee for a useful and informative report.

\end{document}